\documentclass[final,5p,times]{elsarticle} 
\usepackage[hyphens]{url}
\usepackage{todonotes}
\usepackage{graphicx}
\usepackage{mwe}
\usepackage{bm}
\usepackage{amsmath}
\usepackage{amssymb}
\usepackage{tikz}
\graphicspath{{figures/}}
\usepackage[british]{babel}
\usepackage{hhline}
\usepackage{multirow}
\usepackage{subcaption}
\usepackage{bigfoot}
\usepackage{listings}
\usepackage{caption,subcaption}
\definecolor{codegreen}{rgb}{0,0.6,0}
\definecolor{codegray}{rgb}{0.5,0.5,0.5}
\definecolor{codepurple}{rgb}{0.58,0,0.82}
\definecolor{backcolour}{rgb}{0.95,0.95,0.92}

\usetikzlibrary{shapes,arrows}
\lstdefinestyle{mystyle}{
  backgroundcolor=\color{backcolour},   commentstyle=\color{codegreen},
  keywordstyle=\color{magenta},
  numberstyle=\tiny\color{codegray},
  stringstyle=\color{codepurple},
  basicstyle=\ttfamily\footnotesize,
  breakatwhitespace=false,
  breaklines=true,
  captionpos=b,
  keepspaces=true,
  numbers=left,
  numbersep=5pt,
  showspaces=false,
  showstringspaces=false,
  showtabs=false,
  tabsize=2
}

\tikzstyle{decision} = [diamond, draw, fill=white!20,
    text width=4.5em, text badly centered, node distance=3cm, inner sep=0pt]
\tikzstyle{block} = [rectangle, draw, fill=white!20,
    text width=5em, text centered, rounded corners, minimum height=4em]
\tikzstyle{line} = [draw, -latex']
\tikzstyle{cloud} = [draw, ellipse,fill=white!20, node distance=3cm,
    minimum height=2em]

\begin{document}

\journal{Computer Physics Communications}
\bibliographystyle{model1-num-names}

\title{fmmgen: Automatic Code Generation of Operators for Cartesian Fast Multipole and Barnes-Hut Methods}
\author{Ryan Pepper}
\ead{ryan.pepper@soton.ac.uk}
\address{Faculty of Engineering and Physical Sciences, University of Southampton, University Road, Southampton, SO17 1BJ, United Kingdom}
\author{Hans Fangohr}
\ead{hans.fangohr@xfel.eu}
\address{Faculty of Engineering and Physical Sciences, University of Southampton, University Road, Southampton, SO17 1BJ, United Kingdom}
\address{European XFEL, Holzkoppel 4, 22869 Schenefeld, Germany}

\maketitle

\begin{abstract}
The Barnes-Hut and Fast Multipole Methods are widely utilised methods applied in order to reduce the computational cost of evaluating long range forces in $N$-body simulations. Despite this, applying existing libraries to simple problems with higher order source points, such as dipoles, is not straightforward or efficient because individual libraries are optimised towards specific problems, normally solving for the potential and field of a set of Coulombic particles. In this paper we detail the implementation and testing of a software package, fmmgen, in which the source code for Barnes-Hut and Fast Multipole operator functions for calculating calculate the potential, field or both from arbitrary ordered sources is easily generated through symbolic algebra.
\end{abstract}

\section{Introduction}
Implementations of the Fast Multipole Method (FMM) for the Poisson equation ususally utilise a spherical harmonic expansion of the Green's function, which leads to an irreducible representation. Doing this in a computationally efficient way is non-trivial because calculating the spherical harmonic functions necessary for such an expansion effectively requires making use of recursion relations in order to avoid numerical issues. Cartesian Taylor expansions also form a straightforward basis in which to expand the potential, though the multipole expansions are reducible and so are less computationally efficient than spherical harmonic expansions for the high accuracy regime \cite{Coles2018a}. The expansion of the kernel used in different software packages varies widely across different areas of study; in astrophysics a Cartesian basis is primarily used, especially in the Barnes-Hut (BH) method \cite{Dehnen2000, Dehnen2002}, but in chemical molecular dynamics studies it is more usual to make use of the Spherical Harmonic expansion. Similarly, the method utilised varies enormously, and application specific performance optimisations can be made for, which preclude against code-reuse when attempts are made to apply the methods to other problems.

In the past, an approach based on template metaprogramming in C++ was utilised by Visscher and Apalkov \cite{Visscher2010} to provide efficient recursive implementations of the Cartesian operator functions for point dipole and micromagnetic cell sources. A similar templating approach was used by Wang. et. al. in order to implement Cartesian operators for the FMM as applied to the Boundary Element Method \cite{Wang2015}. This approach, however, is not straightforwardly generalisable to languages which do not support code generation at compile time, such as C and Fortran, and it makes it difficult to apply broad optimisations to the templated code. In most FMM and BH implementations based on the Cartesian expansion, hand-written code is therefore used to implement the various FMM operators; in Dehnen's FalcON code this goes as far as hand implementing vectorisation of the various operators \cite{Dehnen2014}. This is difficult, because beyond several expansion orders it is tedious to ensure code correctness, and the number of terms in the multipole expansion grows with $n$ as $n(n + 1)/2$. Implementing operators by hand in this way also makes it difficult to enable code reuse; for example, implementing both the multipolar BH and FMM in the same code base requires a lot of duplication of code.

Generalising to provide efficient operators for point sources of different orders (i.e. point monopoles, point dipoles, point quadrupoles) is also important for adoption of the method. It is important to note that for many problems which are computationally intractable, reduced order models of systems can be constructed using multipoles, by treating objects with complex internal structures as points with a multipole expansion up to some given order, and these sources can often have non-negligible quadrupole terms. \cite{Gareeva2018}

Symbolic code generation is a technique which has, in recent years, been applied to the generation of functions for the computational solution of ordinary and partial differential equations. The FFC library \cite{Alnaes2009} constructs functions for the evaluation of variational forms for assembling finite-element matrices, and is used as part of both the FEniCs and Firedrake projects \cite{Logg2010, Rathgeber2016}. The OpenSBLI project \cite{Lusher2018, Mudalige2019, Jacobs2017} generates finite-difference stencils in the language of the high-performance OPS library \cite{Reguly2018} from symbolic representations of differential equations, while the Devito \cite{devito-compiler} project achieves similar goals uses symbolic code generation functionality in the SymPy library to generate efficient finite-difference kernels written in C. In the context of the fast multipole method, code generation has previously been utilised by Coles and Masella in order to provide an implementation of the Cartesian basis Fast Multipole Method in the closed source PolarisMD code, for the calculation of the electric potential and field from polarisable atoms in molecular dynamics, \cite{Coles2014a} with this work then being extended to support the use of more efficient operators through detracing techniques introduced by Applequist. \cite{Coles2018a, Applequist1984, Applequist1989, Dehnen2014}

In this paper, we describe the implementation and details of an open-source code generation library, fmmgen, \cite{Pepper2019} which produces a set of operators for the Cartesian BH and FMM methods, and provides OpenMP parallelised example implementations of the methods. We draw attention to how optimisations and simplifications can be enabled at different stages in the code generation to improve performance, and comment on the effectiveness of optimisation strategies. We also discuss the inclusion of the software library in the atomistic spin dynamics software Fidimag \cite{Fidimag3.0} to calculate dipolar fields.

\section{Mathematical Basis}
We begin by showing the mathematical details necessary to construct the FMM and Multipolar BH methods to compute the potential and field the Laplace Equation in a Cartesian basis for source points of arbitrary order (i.e. Monopoles, Dipoles, Quadrupoles, ..., $2^n$-poles. We denote the minimum `order' of point sources in a system source as $s$, such that a monopole has order $s=0$, a dipole $s=1$, etc.

We here use the mathematical notation of monomials, which is widely used in the Fast Multipole literature. Here:

\begin{align*}
\bm{n} &= (n_x, n_y, n_z) \\
\bm{n} + \bm{m} &= (n_x + m_x, n_y + m_y, n_z + m_z) \\
\bm{n}! &= n_x!n_y!n_z! \\
\bm{r}^{\bm{n}} &= x^{n_x} y^{n_y}z^{n_z} \\
\lvert\bm{n}\rvert &= n_x + n_y + n_z \\
\binom{\bm{n}}{\bm{k}} &= \binom{n_x}{k_x}\binom{n_y}{k_y}\binom{n_z}{k_z}
\end{align*}

Consider the expansion of the Coulomb Potential from two well-separated cells $A$ and $B$, with centres $\bm{z}_a$ and $\bm{z}_b$, and containing points $\bm{x}_a$ and $\bm{x}_b$ respectively. We define vectors $\bm{r}_a = \bm{x}_a - \bm{z}_A$ and $\bm{r}_b = \bm{x}_b - \bm{z}_{B}$. When a charge $q_a$ is located at $\bm{x}_a$, the potential at $\bm{x}_b$ can be evaluated as:

\begin{align}
\phi(\bm{r}) = \frac{q}{\lvert \bm{x}_b - \bm{x}_a \rvert}
\end{align}

\noindent
Taylor expanding this around the point $x_a$ and truncating at order $p$ gives an approximate function for the evaluation of the potential:

\begin{align}
\phi{(\bm{x}_b - \bm{x}_a)} &\approx q_a \sum_{\lvert \bm{n} \rvert = 0}^p \frac{(-1)^{\bm{n}}}{\bm{n}!}(\bm{x}_a - \bm{z}_a)^{\bm{n}} \nabla^{\bm{n}} \phi(\bm{x}_b - \bm{z}_a)
\label{eq:single-expanded}
\end{align}

\noindent
By grouping terms, a multipole term defined around the centre $\bm{z}_A$ can be written:\footnote{This definition varies between fields and authors. Notably, the factor of $(-1)^{\bm{n}}/\bm{n}!$ is often absorbed into the local expansion definition. It is also worth noting that the definition of the dipole and quadrupole moments can vary; for e.g. in Chemistry the dipole moment vector for a two charge system is normally given as directed from positive to negative charge; in Physics this is reversed.}
\begin{equation}
\mathcal{M}_{\bm{n}}(\bm{z}_A) = \frac{(-1)^{\lvert\bm{n}\rvert}}{\bm{n}!}q_a (\bm{x}_a - \bm{z}_a)^{\bm{n}}
\label{eq:coulomb-multipole}
\end{equation}
\noindent
For a given $\mathcal{M}_{\bm{n}}$ term centred at $\bm{z}_a$, the shifted multipole expansion at a centre $\bm{z}_a'$ can be derived through the substitution of $(\bm{x}_a - \bm{z}_a) = ((\bm{x}_a - \bm{z}_a') + (\bm{z}_a' - \bm{z}_a))$, expanding out in powers and substituting multipole terms where recognised.
\begin{equation}
\mathcal{M}_{\bm{n}}(\bm{z}_a') =
\sum_{\lvert \bm{k}\rvert = 0}^{p - \lvert\bm{n}\rvert}
 \frac{(\bm{z}_a - \bm{z}_a')^{\bm{k}}}{\bm{k}!} \mathcal{M}_{\bm{n} - \bm{k}}(\bm{z_a})
\label{eq:multipole-shift}
\end{equation}
Using (\ref{eq:multipole-shift}), expressions for calculating the multipole expansion of arbitrary order source particles can be written by considering a `source' multipole $\mathcal{S}_{n}$. For a Coulomb charge, such that $\mathcal{S}_{(0, 0, 0)} = q$, and all other terms would be zero. For a dipole, $\mathcal{S}_{(1, 0, 0)} = \mu_x$, $\mathcal{S}_{(0, 1, 0)} = \mu_y$ and $\mathcal{S}_{(0, 0, 1)} = \mu_z$, with all other terms zero. Mixed systems can also be considered. Thus, in an arbitrary system where the lowest order of source is $s$, the expansion can be written:
\begin{equation}
\bm{M}_{\bm{n}}(z) = \sum_{\lvert \bm{k} \rvert = 0}^{p - \lvert \bm{n}\rvert} \frac{(\bm{z}_a - \bm{x}_a)^{\bm{k}}}{\bm{k}!}\mathcal{S}_{\bm{n}-\bm{k}}
\label{eq:source-shift}
\end{equation}
For the charge only case, we see that we can straightforwardly recover through Eq \ref{eq:coulomb-multipole} through the knowledge that all terms except $S_{(0,0,0)}$ are zero.

The potential can then be rewritten in terms of these Multipole terms. This expression forms the basis of the multipolar Barnes-Hut method.
\begin{equation}
\phi{(\bm{x}_B - \bm{x}_A)} \approx \sum_{\bm{n} = s}^{p} \frac{(-1)^{\bm{n}}}{\bm{n}!} \bm{r}_a^{\bm{n}}\mathcal{M}_{\bm{n}} \nabla^{\bm{n}} \phi{(\bm{x}_B - \bm{z}_A)}
\end{equation}
Taking a further expansion, this time around $\bm{z}_B$, and truncating such that the maximum order of terms is the same gives:
\begin{equation}
\phi{(\bm{x}_B - \bm{x}_A)} \approx \sum_{\bm{n} = s}^{p} \sum_{\bm{m}=0}^{p-\lvert\bm{n}\rvert - s} \frac{(-1)^{\bm{n}}}{\bm{n}!\bm{m}!} \bm{r}_b^{\bm{m}} \mathcal{M}_{\bm{n}} \nabla^{\bm{n} + \bm{m}} \phi{(\bm{x}_B - \bm{z}_A)}
\label{eq:double-expanded}
\end{equation}
Grouping terms again in Eq. \ref{eq:double-expanded}, a local expansion can be evaluated centred around $\bm{z}_B$.
\begin{equation}
\mathcal{L}_{\bm{n}}(\bm{z}_B) = \sum_{\lvert \bm{m} \rvert = 0}^{p-\lvert\bm{n}\rvert - s} \frac{(-1)^{\bm{n}}}{\bm{m!}}\mathcal{M}_{\bm{m}}(\bm{z}_A)\nabla^{\bm{n + m}} \phi(\bm{z}_B - \bm{z}_A)
\end{equation}
Then, the potential can be evaluated in terms of the local expansion.
\begin{equation}
\phi{(\bm{x}_B)} \approx \sum_{\lvert\bm{n}\rvert=s}^{p} \frac{1}{\bm{n}!}(\bm{x}_b - \bm{z}_b)^{\bm{n}} \mathcal{L}_{\bm{n}}(\bm{z}_B)
\end{equation}
Derivatives of the potential then be calculated by differentiating this expression with respect to the component axis:
\begin{equation}
\frac{\partial^{\bm{k}}\phi}{\partial \mathbf{r}^{\bm{k}}}  \approx \sum_{\lvert\bm{n}\rvert=s + \lvert \bm{k} \rvert}^{p} \frac{1}{(\bm{n} - \bm{k})!}(\bm{x}_b - \bm{z}_b)^{\bm{n} - \bm{k}} \mathcal{L}_{\bm{n}}(\bm{z}_B)
\end{equation}
If the order of the derivative is greater than $p - s$, this expression is not sufficient. In this case, a finite-difference approximation must be used.

\section{Implementation}
\subsection{Operator Generation}
Here, we attempt to give a description of the open source code generation framework, fmmgen, \cite{Pepper2019} is implemented and how it can be used. The framework is built in Python, using the symbolic algebra package SymPy \cite{Meurer2017}, and generates source code output in C and C++, with the reasoning that code generated in these languages by the framework can be straightforwardly incorporated into other projects without great difficulty or the requirement of large dependencies.

The code generation of each of the multipole operator equations can be broken up into different stages, each of which can be used independently. The user must specify the minimum source order $s$, the maximum expansion order $p$, and the output they desire (potential, field, or both). From these parameters, a mapping between $\bm{n}$ values and one-dimensional array indices is created. By default this mapping is lexicographic, i.e. ((0,0,0), (1,0,0), (0,1,0), (0,0,1), (2,0,0), (1,1,0), ...), such that the total monomial order of a given term is strictly increasing. Nonetheless, if another ordering is preferred (for e.g. in some fields the quadrupole moments are ordered differently), it is possible to use change this by simply using a different array mapping. If it is known in advance that certain terms will always be zero, terms can be removed from the mapping in order to create simpler symbolic representations of the multipole and local expansion operators. We also make use of the source order parameter given by the user to reduce the memory needed to store the multipole and local expansions; this is possible because it is not possible to construct a multipole with a net $n^{th}$-moment from sources of are of order $s > n$.

A set of expansion functions are implemented for the Fast Multipole Method, which are used to construct symbolic representations of $\mathcal{M}_{\bm{n}}$, the Particle-to-Multipole (P2M) operator, and $\mathcal{L}_{\bm{n}}$, the Multipole-to-Local operator,  at a given $\bm{n}$, as well as the shifting operators for these, the Multipole-to-Multipole (M2M) and Local-to-Local (L2L) operators. These functions must make reference to the mapping, in order to return the correct array indices. Generator functions use the set of expansion functions and iterate through the full list of $n$ values needed for a particular problem, and an array representation of each operator is formed. This is repeated for each expansion order, and a least-recently-used (LRU) cache is used in the generation stage to reduce the code generation time.
We finally generate a symbolic representation of the operator functions for both the Barnes-Hut and Fast Multipole Method which can calculate the required quantities from a multipole (M2P) or local expansion (L2P), or from another source (P2P).

Once the full set of symbolic operators is generated, a code writing class is used to turn the symbolic representation of the operators into C or C++ code. While the SymPy library can provide some basic code-generation functionality, by default it generates unoptimised code which leaves much room for improvement in performance terms. To this end, we implemented a set of optimisations which can be enabled and disabled at the code generation stage by the user of the library. We leave these as options rather than enabling by default, because it is then easy to test that the optimisations affect only the performance, and because the optimised code is often more difficult to read and hence debug.

Coles et. al. previously discussed how in code generation of multipole operators, \cite{Coles2014a} they reduce the number of mathematical operations in the code through Common Subexpression Elimination (CSE), which analyses the code for repeated calculations across multiple lines, and pulls these out as factors. Prior to using CSE, we preprocess the operators to increase the chance of finding common subexpressions. These preprocessing stages rationalise powers (for e.g. replacing ($x^2)^2$ with $x^4$, factor terms, and remove extraneous multiplications which sometimes appear in the code generation stage (e.g. $(1.0)x$). In Figure \ref{fig:cse}, we show the effect that CSE has on the Multipole-to-Local operator.

\begin{figure*}[!ht]
  \centering
  \begin{subfigure}[b]{\textwidth}
    \begin{lstlisting}[language=C++]
    void M2L_1(double x, double y, double z, double * M, double * L) {
        double R = sqrt(x*x + y*y + z*z);
        double D[4];
        D[0] = (1 / (R));
        D[1] = -1.0*x/(R*R*R);
        D[2] = -1.0*y/(R*R*R);
        D[3] = -1.0*z/(R*R*R);
        L[0] += D[0]*M[0] + D[1]*M[1] + D[2]*M[2] + D[3]*M[3];
        L[1] += D[1]*M[0];
        L[2] += D[2]*M[0];
        L[3] += D[3]*M[0];
    }
    \end{lstlisting}
    \caption{M2L Operator without CSE}
    \label{labelname 1}
  \end{subfigure}
  \begin{subfigure}[b]{\textwidth}
    \begin{lstlisting}[language=C++]
    void M2L_1(double x, double y, double z, double * M, double * L) {
        double Rinv = pow(x*x + y*y + z*z, -0.5);
        double D[4];
        double Dtmp0 = (Rinv*Rinv*Rinv);
        D[0] = Rinv;
        D[1] = -Dtmp0*x;
        D[2] = -Dtmp0*y;
        D[3] = -Dtmp0*z;
        L[0] += D[0]*M[0] + D[1]*M[1] + D[2]*M[2] + D[3]*M[3];
        L[1] += D[1]*M[0];
        L[2] += D[2]*M[0];
        L[3] += D[3]*M[0];
    }
    \end{lstlisting}
    \caption{M2L Operator with CSE}
    \label{labelname 2}
  \end{subfigure}
  \caption{Here, we see the affect of enabling common-subexpression elimination for the 1st Order Multipole-to-Local operator in the FMM method when $s = 0$. Prior to enabling this subexpressions such as $1/R^3$ are repeated multiple times across multiple lines of code as in (a), while with it enabled, these are factored out into temporary stack variables as in (b).}
  \label{fig:cse-code}
\end{figure*}

The optimisations have the greatest effect on performance in the calculation of the Multipole-to-Local operator for the FMM and Multipole-to-Particle operators for the Barnes-Hut method, which make use of the calculation of the derivatives of $1/r$ up to a given order.

In traditional codes, the computations of derivatives of these derivatives up to an order $p$ are usually performed incrementally, such as by using the $\mathcal{O}(p^6)$ formula of Cipriani and Silvi \cite{Cipriani1982} or using an $\mathcal{O}(p^4)$ recursive formula as described by Challacombe et. al. \cite{Challacombe1995}. With the code-generation, we were able to implement straightforwardly an optimisation noted by Dehnen \cite{Dehnen2014} by making use of the harmonicity of the Poisson Green's function, which allows us to calculate derivatives as:

\begin{equation}
\nabla^{\bm{n} + (0, 0, 2)}\phi = - \nabla^{\bm{n} + (2, 0, 0)}\phi - \nabla^{\bm{n} + (0, 2, 0)}\phi
\end{equation}

This reduces the number of mathematical operations for higher order calculations. We do note however, that while the SymPy library provides some metrics for the number of mathematical operations in given expressions, these are not an effective way of deducing the computational cost of generated code, because the choice of compiler and the enablement of compiler optimisations drastically affects the FLOP count, and because some operations take more clock cycles than others. This means that for accurate FLOP counts, tools such as Intel VTune must be used at runtime.

The code also supports the replacement of evaluations of \verb'pow(x, n)' (or \verb'std::pow(x, n)' in C++), where $n$ is a positive or negative integer value, with multiplication. It is well known that this can be an effective optimisation in numerical codes, but in practice it can be tedious to implement, and beyond a certain point round off errors begin to accumulate. \cite{Wicht2017} In the code generation stage, these operations can be replaced up to some maximum $n_\text{max}$, the optimum which can be determined through profiling for a given architecture, precision and compiler combination.

\subsection{Methods}
We implemented both the Barnes-Hut and Fast Multipole Method in a single code base using the generated operator functions. In this, we make use of an octree data structure, whereby the simulation domain is recursively subdivided into octants depending on the particle density, controlled by a parameter $n_\text{crit}$, which controls the maximum number of particles in an octant before it is split. The implementation of our octree structure is such that the memory comprising of the multipole and local expansion arrays is contiguous, to allow better cache coherency. Unlike in some codes, we use the cell centre as the expansion centre; while for gravitational systems the centre of mass is an obvious choice as the dipole term in a cell will vanish, for higher order sources and mixed systems, the choice is not so obvious. For the BH method, we evaluate the multipole expansion on cells at the lowest level of the tree, and then pass this upwards using the M2M operators. Then, for each particle, located at $\bm{x}_p$ the tree is traversed from the top level downwards. A cell is considered to be near to a particle if it meets the Barnes-Hut multipole acceptance criterion:
\begin{equation}
\frac{r_{cell}}{\lvert x_p - x_c\rvert} < \theta_{\text{BH}}
\end{equation}
which relates the cell size to the distance, and an opening angle parameter $\theta$, which is a user supplied parameter which controls the accuracy.

If a cell has no child cells, and the cell does not meet the acceptance criterion, then the cell's particles are looped through, and the interaction is calculated directly using the Particle-to-Particle (P2P) operator. If the criterion is met, then the interaction between the cell and the particle is instead computed using the Multipole-to-Particle (M2P) operator. Finally, if the cell has child cells, then the procedure is repeated on these.

For the FMM, we implemented the dual-tree traversal algorithm which has seen widespread adoption, rather than the classic FMM introduced by Greengard and Rokhlin in which cell-cell interactions only occur between neighbouring cells and their children, \cite{Yokota2013} as this this has much in common with the Barnes-Hut approach. The initial procedure here is the same as the Barnes-Hut method; multipoles are computed for cells on the lowest level of the tree and then shifted upwards. Then, the tree is traversed from top to bottom. Cells which fulfill the multipole acceptance criterion:
\begin{equation}
\frac{r_{c_A} + r_{c_B}}{R} < \theta_{\text{FMM}}
\end{equation}
interact via the Multipole-to-Local (M2L) operator, while cells which do not are recursed into until either their children fulfil the criteria, or a leaf cell is reached, at which point the cells interact directly. For more straightforward parallelisation, as opposed to the task-based parallelism favoured by some authors, we traverse the tree at initiation in our test implementation, and store the sorted interaction lists which can then be iterated through with loop-based parallelism.

\section{Testing}
\begin{figure*}[ht]
  \centering
  \includegraphics[width=6.0in]{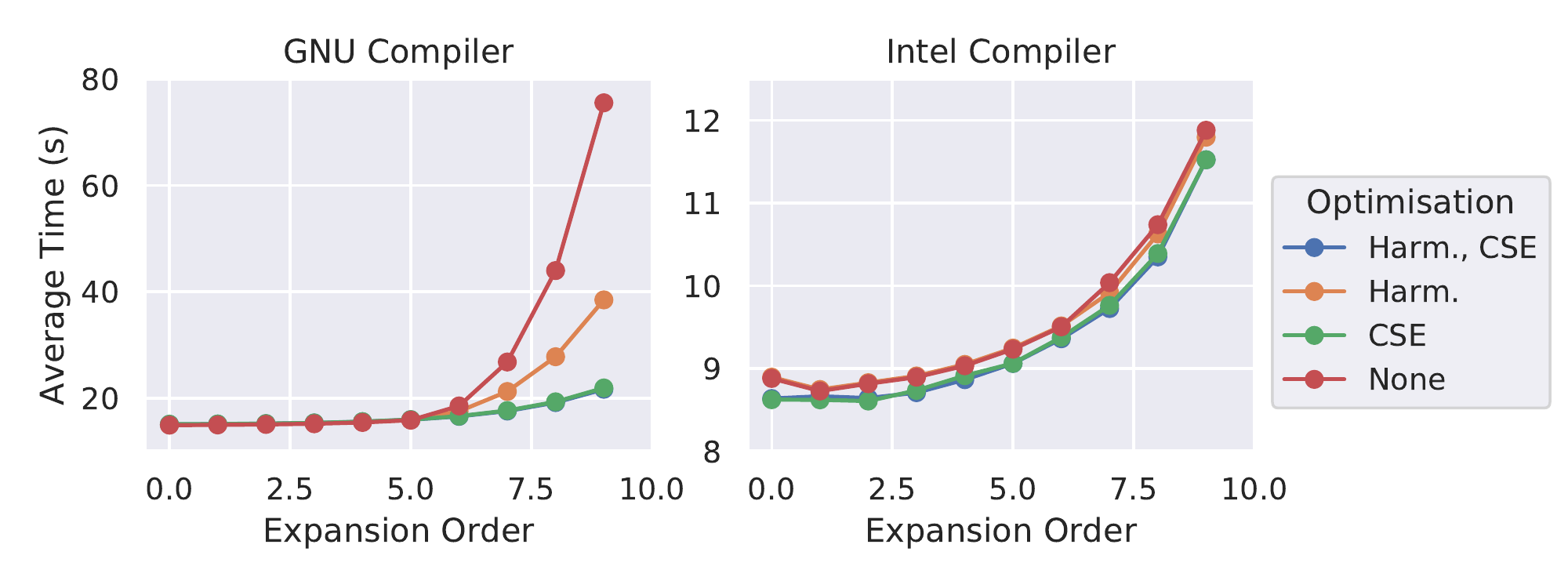}
  \caption{Performance between compilers and with CSE and Harmonic Derivative evaluation disabled and enabled. We see a much greater impact of the code generation optimisations using the GNU g++ compiler than with the Intel Compiler.}
  \label{fig:cse}
\end{figure*}
We provide a test application with the library which can be configured to allow the evaluation of the potential and/or field from a set of source particles of arbitrary order, using either the Barnes-Hut or FMM approach, which allows for a straightforward comparison between the two methods and their performance. We ran tests with this text executable on a machine with a 4-core 3.4GHz Intel i7 6700 machine. We note that this processor is affected by the Spectre and Meltdown vulnerabilities, and testing was performed with the Linux kernel version 4.15.0-55-generic, which includes mitigations for this, which have been reported to affect the performance of some HPC applications. \cite{Simakov2018} The executables were compiled with both the the Intel and GNU g++ compilers to allow comparison between the performance.\footnote{The executables were compiled with the Intel Compiler v.19.0.3.199 from Parallel Studio 2019 Update 3 and g++ v.7.5.0.} Tests were performed with OpenMP enabled, and with options set to prevent thread migration between cores and idle threads from sleeping, and with hyperthreading disabled. All of the timing results shown below are averaged over three runs in order to reduce the effect of system calls and background processes on the runtime measurement.

Initially, we tested how the performance optimisations described in the previous section affected the performance of the potential and field calculation via the Fast Multipole Method, for a system of $10^5$ randomly distributed charged particles in $[-10^{-9}, 10^{-9}]^3$, with fixed values of $\theta = 0.3$ and $n_\text{crit}=128$. We compiled executables for both compilers with generated operators with CSE and the computation of derivatives through the reuse of results and the harmonicity property enabled and disabled, the results of which are shown in Figure \ref{fig:cse-code}. We found that in general, the timing results were relatively consistent, with the runtime increasing progressively with the expansion order. With the GNU compiler, enabling the harmonic derivatives optimisation led to a decrease in performance at $9^\text{th}$ order of around $50\%$ while enabling CSE led to around a $75\%$ decrease. For the Intel compiler, the corresponding decreases were around $1\%$ and $2\%$. At lower expansion orders, we see very little performance increase, and this is because there are fewer opportunities for eliminating common factors in expressions. The reason behind the difference between the GNU and Intel compilers was investigated. Analysis with Intel VTune showed that substantial numbers of the repeated operations at high optimisation levels were cached in compilation of the non-CSE enabled code with the Intel Compiler, but not with the GNU compiler. In both cases, there was only a marginal difference in performance when both optimisations were enabled. All subsequent tests to this were run with the Intel executable with both CSE and Harmonic derivatives enabled.

\begin{figure*}[h!]
  \centering
  \includegraphics[width=6.0in]{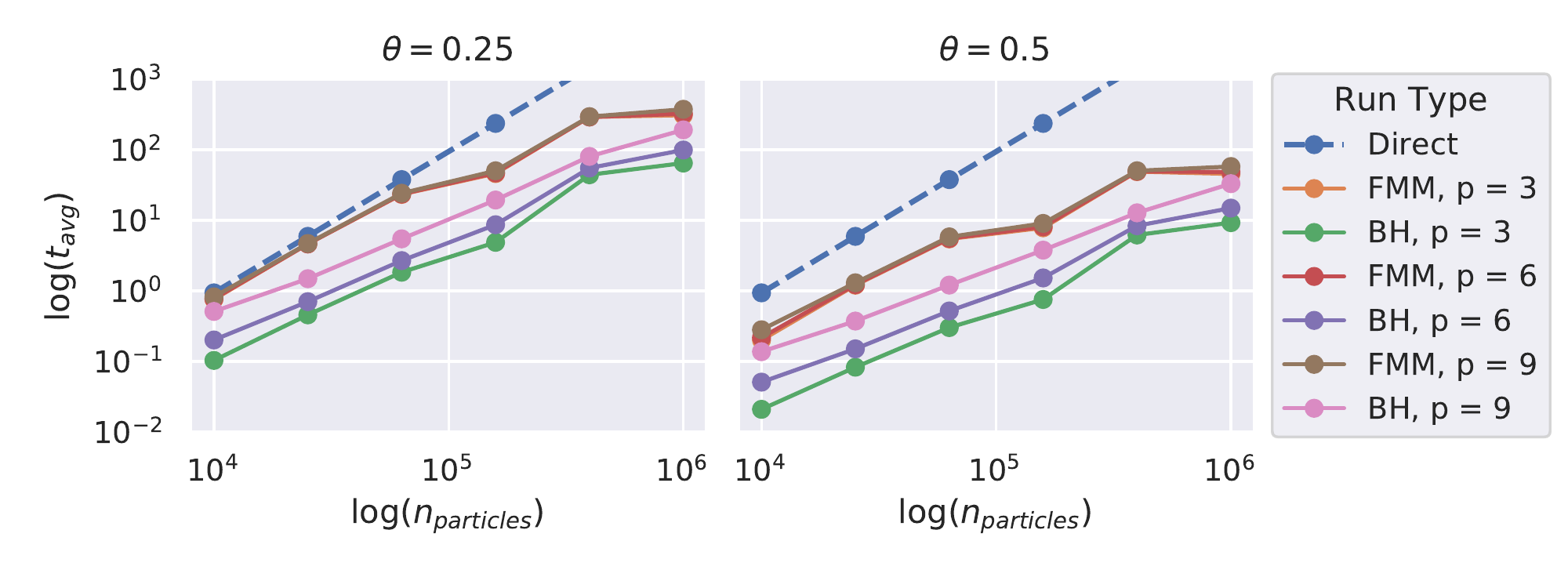}
  \caption{We show how performance varies when varying the opening angle parameter $\theta$ and the expansion order $p$ for both the Barnes-Hut and FMM methods. We note that the run time is much less affected by increasing expansion order for the FMM method compared to the Barnes-Hut method.}
  \label{fig:cse}
\end{figure*}
In Figure 2, we show the scaling of the FMM and BH methods with regards to the number of particles, where the number of particles is chosen such that they are equidistant in log-space. We can see that for numbers of particles up to $10^6$, the BH method outperforms the FMM. In both cases, the exact breakeven over the direct method depends on the expansion order, but is less than 1000 charges. We can see that increasing the expansion order gives a clear delineation of the runtime of the Barnes-Hut method while in the FMM, there is less of an impact; this is because the M2P kernel is evaluated many more times in the BH method than the equivalent M2L kernel is in the FMM method, and it is why the method scales more poorly ($\mathcal{O}(n \log n)$ for BH vs $\mathcal{O}(n)$ for the FMM) at very large numbers of particles.

We note that the two multipole acceptance criterion are not directly equivalent, despite having a similar controlling effect on accuracy, because in the BH it directly relates the particle distance to a cell and it's size, while in the FMM it is a cell-cell parameter. As a result of this, to achieve similar error characteristics with the two methods, $\theta_\text{FMM}$ should be around twice $\theta_{BH}$. This can be seen in Figure 3, where we show the error distributions for the two methods at different expansion order at $\theta = 0.25, 0.5$ for a system of $50000$ particles with $n_crit = 128$. Here, we can see that the

\begin{figure*}[h!]
  \centering
  \includegraphics[width=5.0in]{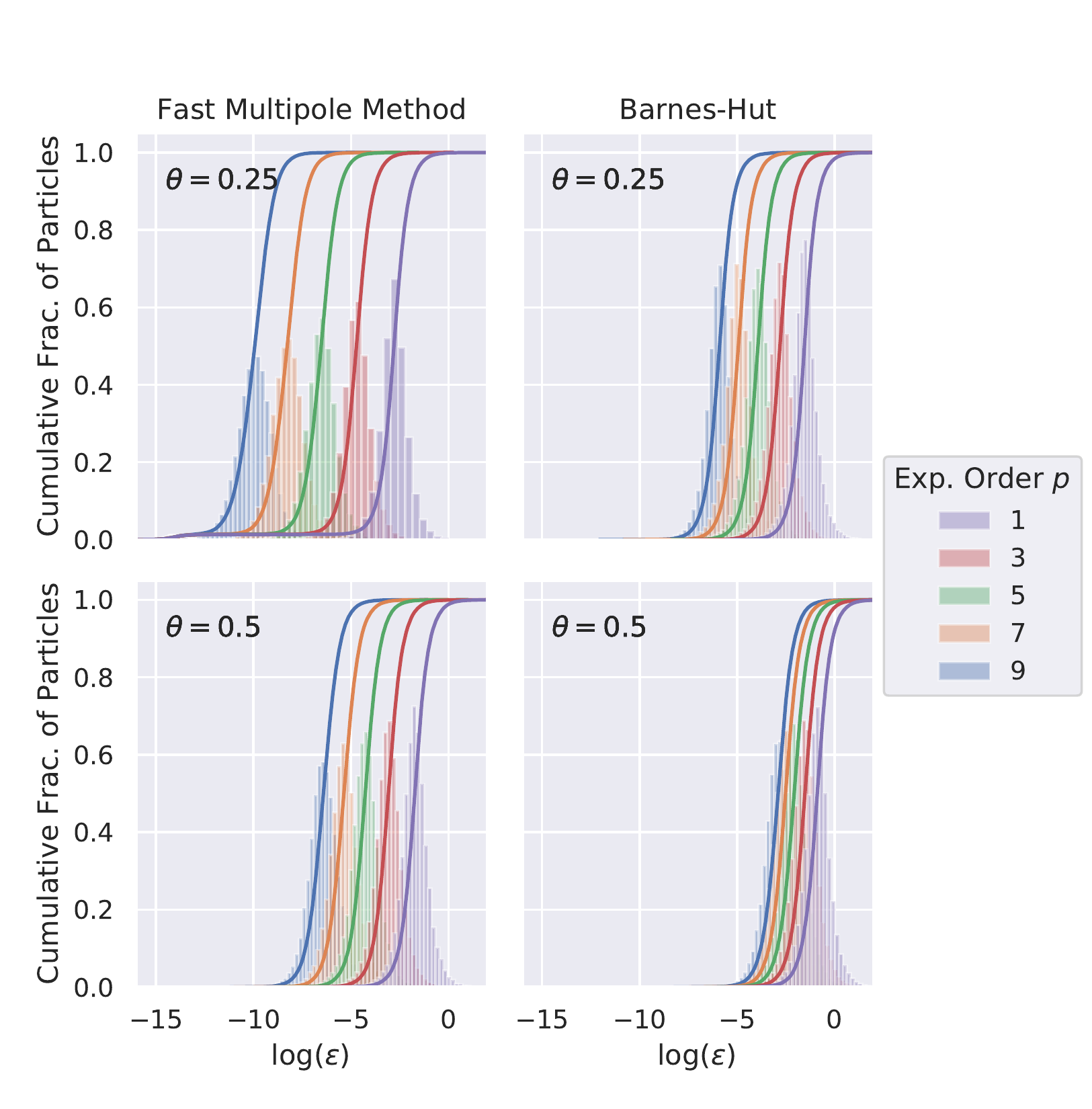}
  \caption{Histogram and cumulative distribution of error for Coloumb potential calculation for the Fast Multipole and Barnes-Hut methods. With increasing expansion order, we see that the median error decreases.}
  \label{fig:error}
\end{figure*}

\begin{figure*}
  \centering
  \includegraphics[width=4.5in]{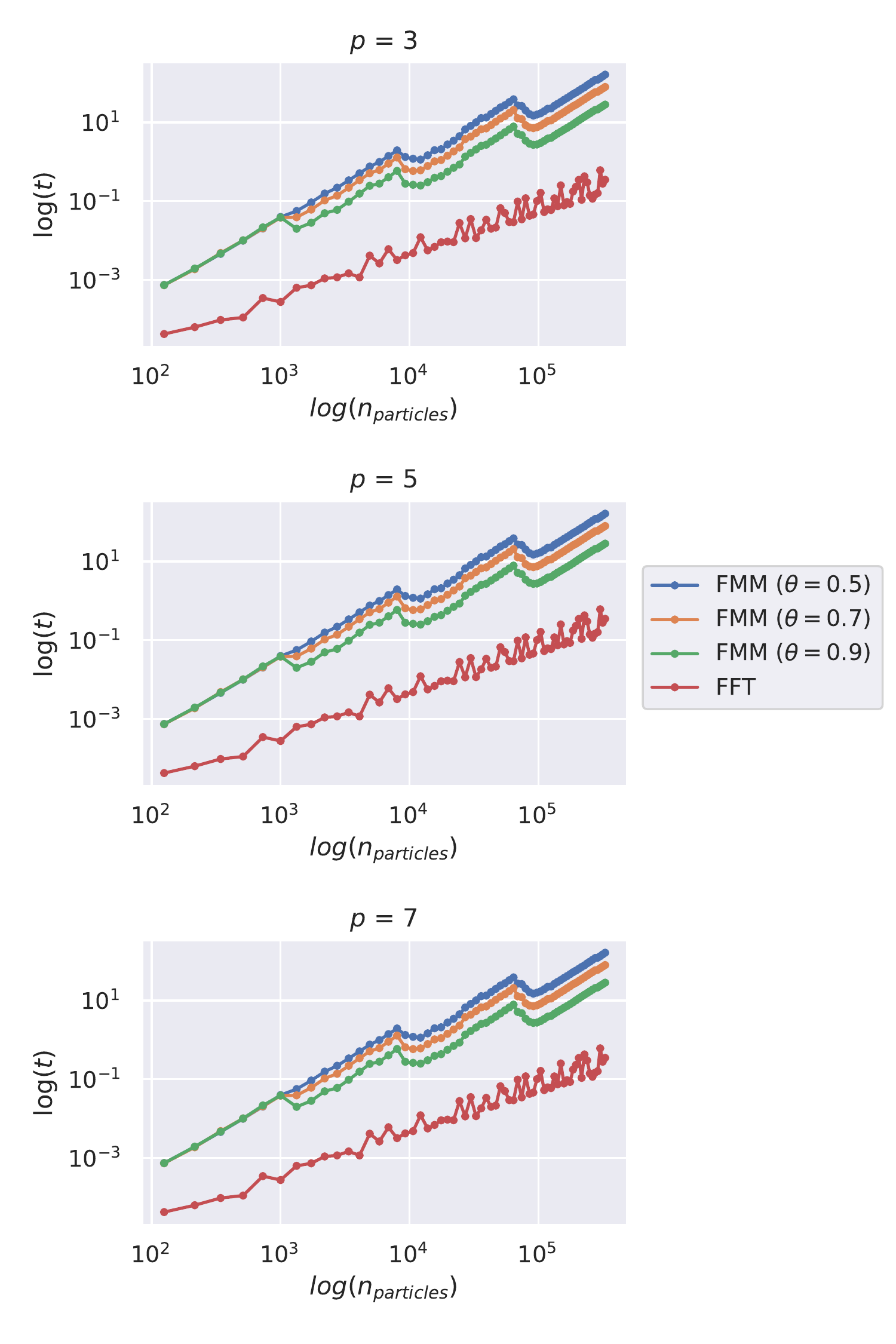}
  \caption{Here we show the performance of the FMM against the FFT for dipolar field calculation, by comparing runtime as the number of spins in a cube is increased at different expansion orders. We note that between expansion orders, we see a negligibly small difference in performance with variation in $\theta$ having much more of an impact on runtime.}
  \label{fig:fmm-fft}
\end{figure*}

As a real-world test case, we integrated the dipole field FMM calculation generated by the library into the atomistic spin dynamics component of the computational nanomagnetism software Fidimag \cite{Bisotti2018, Fidimag3.0}. To check the implementation, we compared it against the standard technique in this field, which is to use a Fast Fourier Transform (FFT) accelerated convolution in order to sum the field contributions from dipoles placed on a lattice \cite{Mansuripur1988, Yuan1992, Nowak2007}. We constructed a test case comprising of a system of atomic dipoles in a cubic arrangement with $n$ spins on each axis resulting in $n^3$ spins, and varied $n$ between $5$ and $70$. We computed the field with the convolution method using the Fast Fourier Transforms computed from the library FFTW with OpenMP parallelism enabled. In order that the comparison was fair, we neglected any start up time which is one-off, and so do not include the pre-computation of the demagnetising tensor for the FFT technique or the tree construction for the FMM technique.

We show the performance results in Fig. \ref{fig:fmm-fft}. In all tests, we found that the FMM method was around an order of magnitude worse in performance terms compared to the FFT convolution technique. We also note that for some $\theta$ values ($\theta > 0.7$), in realistic test simulations in which the Landau-Lifshitz-Gilbert equation was used to relax the system, we found that simulations either failed to converge using the FMM, or took more integration steps to do so, as a result of the loss of accuracy in the method. This suggests that, at least on parallel shared memory architectures, using the FMM for dipolar field calculations, at least as implemented here, is not an effective method method for lattice simulations.

Despite this, the inclusion of the FMM method into our code Fidimag is designed such that it enables the study of systems where particles do not lie on a lattice, enabling the computation of the dipolar field in problems where it was not previously possible. We note that the atomistic spin dynamics codes Vampire \cite{Evans2014} and Vinamax \cite{Leliaert2015} make use of approximations for computing the dipole field that are similar to the Barnes-Hut method with $p = 1$ and $s = 1$. From our own tests, we found that approximation at this level of expansion order is not sufficient to maintain an acceptable level of accuracy in simulations in general, because it can lead to an error on the dipolar field of over $100\%$ on individual particles - we note that for example, in Fig. \ref{fig:error}, we can see that the error distributions have a substantial tail, necessitating the use of high expansion orders to put a reasonable constraint on the error of the field at individual points where the potential is calculated. The effect of such large errors may or may not manifest itself in simulations, and is strongly dependent on other parameters and the relative strength of the dipolar field against other energy terms.

Our results contradict prior performance studies on the fast multipole method in atomistic lattice systems, where the method showed speed-ups over the FFT convolution method for the numbers of particles commonly used in atomistic simulations. We note that the method shown in one paper promising speed-ups from the Cartesian FMM used the scalar non-parallelised FFT routine from Numerical Recipes \cite{numrecipes}, which was likely to be significantly slower than the FFT methods in FFTW (originally released in 1999, and with the much improved version 3 released in 2003 which added vectorised forms of the FFT) even at the time of publication \cite{Zhang2009a}. Though we have chosen here to show the results by way of comparison with the FFT in FFTW due to this being freely available across architectures and operating systems, we note that performance of the FFT through the FFTW interface supplied in Intel's Math Kernel Library was found to be around 2.5x that of FFTW on the same hardware used in this study, and so the FMM fared worse by comparison under these circumstances.

\section{Discussion}
In this work, we have implemented and shown the efficacy of code generation for the Multipolar Barnes-Hut and Fast Multipole Methods, and have described the implementation of this into an publicly available framework. While we have achieved substantial increases in performance over the direct method, there are several areas in which further progress can be made. Notably, the use of an irreducible representation of the operator functions through the use of a detracing operator can reduce the storage space needed for the Cartesian FMM. \cite{Coles2018a} In addition, while we have not yet attempted to apply explicit vectorisation in the code generation stage, the formulation of the code generation system means that doing so across the whole code base should be a more straightforward exercise than a hand written version.

It is important to note that while the algebraic complexity of the spherical harmonic expansion is lower (at $\mathcal{O}(p^4)$ for a naive implementation or $\mathcal{O}(p^3)$ when rotations are used to reduce the cost of the local expansion translation), at low orders it has been shown by various authors that the computational cost of using the Cartesian method is often still lower. It has, however, been shown by the proliferation of consumer-grade GPU hardware in computational research that in many cases, accuracy of less than $10^{-7}$ is sufficient in many numerical applications. It is with this in mind that there is still much to recommend about the Cartesian approach over the Spherical Harmonics technique.

It is of our opinion that the specialised nature of many fast multipole libraries towards specific problems means that heirarchical methods have not been as successfully adopted as other numerical techniques, and indeed, part of our own motivation for this work was in the difficulty of applying existing packages to our own problems of interest, namely nanomagnetic dipoles. We note that, for example, in gravitational systems where the domain origin is chosen as the centre of mass, the dipole moment will always vanish. \cite{Dehnen2002, Coles2018a}. A specific and widely used optimisation for the fast multipole method in this case, therefore, is to neglect entirely the calculation of the dipole moments in a system, which precludes the reuse of a hand-written gravitational FMM code for other applications where the dipole moments are non-vanishing, without some modification.

\section{Data Access}
The fmmgen software is hosted at \url{https://github.com/rpep/fmmgen}, and an archive copy of the version used to perform this study is stored on Zenodo at \url{https://doi.org/10.5281/zenodo.3842591}. The code, data and scripts are available to reproduce the study and figures and are hosted publicly on Zenodo at \url{http://doi.org/10.5281/zenodo.3842584}.

\section{Acknowledgements}
This work was financially supported by EPSRC Doctoral Training Centre Grant EP/L015382/1. We thank J. Coles from Technische Universit\"at M\"unchen for his helpful suggestions on testing the correctness of the implemented operator functions.

\onecolumn{
\section{References}
\bibliography{library}
}
\end{document}